\documentclass[aps,preprint]{revtex4}
\usepackage{graphicx,amsmath}

\draft
\begin{document}
\author{Min He$^{a}$, Jian-Feng Li$^{a}$, Wei-Min Sun$^{a,b}$, and Hong-Shi Zong$^{a,b}$}
\address{$^{a}$ Department of Physics, Nanjing University, Nanjing 210093, P. R. China}
\address{$^{b}$ Joint Center for Particle, Nuclear Physics and Cosmology, Nanjing 210093, China}
\title{Quark number susceptibility around the critical end point}
\begin{abstract}
The quark number susceptibility is expressed as an integral in terms
of dressed quark propagator and dressed vector vertex. It is then
investigated with the latter two- and three-point functions
confronted with a Dyson-Schwinger equation model which accommodates
both finite temperature and baryon chemical potential. The critical
end point in the phase diagram is identified and the behavior of the
quark number susceptibility around the critical end point is
highlighted. The qualitative features found agree
with recent lattice QCD simulation results.

\bigskip


\bigskip
 E-mail: zonghs@chenwang.nju.edu.cn.

\bigskip


\end{abstract}
\maketitle
\section{Introduction}
Underlying our understanding of strong interaction, Quantum
Chromodynamics (QCD) is remarkable in that its fundamental degrees of
freedom -- quarks and gluons are confined and the (approximate)
chiral symmetry in the light quark sector is dynamically broken. On
the other hand, at sufficiently high temperatures and/or baryon chemical
potentials, QCD is believed to undergo a phase transition into the
deconfined quark gluon plasma (QGP) (for a recent review, see, e.g.,
\cite{a1}), and probably at the same time, the spontaneously broken
chiral symmetry gets restored \cite{a2}. It is the main task of the
on-going BNL-RHIC and one of the goals of CERN-LHC to create and
probe such a new state of QCD matter.

The phase transition of confined hadronic matter to QGP at finite
temperature and vanishing baryon chemical potential has been
extensively studied by lattice QCD simulations as well as
phenomenological models. The chiral phase transition of QCD with two
flavors in the chiral limit is likely to be of second order
\cite{a3} and turns to a smooth crossover in the case of physical
finite current quark masses \cite{a4,a5,a6}. At finite chemical
potentials and lower temperatures, more and more robust lattice
simulations indicate that QCD experiences a first-order phase
transition \cite{a7,a8}. Such a scenario of QCD phase transition
suggests that there should exist a critical end point (CEP) on the
phase diagram -- the point where the first-order phase transition
line terminates (for a recent review on CEP, and especially the
argument that CEP must exist, see \cite{a9}).

The search for the QCD CEP has attracted considerable attention over
the years from both theoretical and experimental aspects
\cite{a10,a11,a12,a13,a14,a15,a16}. Note that the end point of a
first-order phase transition line is a critical point of the second
order -- in this connection, the water liquid-vapor phase transition
makes a more familiar example. This means that the CEP would be
characterized by enhanced long wave-length fluctuations which lead
to singularity in thermodynamic observables. The authors of
Ref. \cite{a10} suggest that quark number susceptibility should
develop some singularity near the CEP and its divergence could be a
signal of the CEP. From then on, a lot of phenomenological model
investigations \cite{a17,a18,a19,a20,a21,a22} as well as lattice
simulations \cite{a23,a24,a25} toward this susceptibility have been
attempted, all aimed at determining the location of the CEP on the
temperature ($T$)-chemical potential ($\mu$) plane. On the
experimental side, recent plans for verification of the CEP in heavy
ion collisions have concentrated on energy scans to access as broad
as possible range of $T$ and $\mu$ values necessary for the search
for the CEP. It is argued that the range of $T$ and $\mu$ values for
the CEP search can be accessible by combined results from energy
scans at FAIR, SPS and RHIC \cite{a16}. Also notable is that, apart
from susceptibilities, transport coefficients, like the ratio of
viscosity to entropy density ($\eta/s$) can be other probes for the
CEP \cite{a15,a16,a26}.

In the present work, we extend our previous continuum study of quark
number susceptibility \cite{a3} to finite chemical potential in a
Dyson-Schwinger equation model and aim at locating the CEP. We first
derive a closed integral expression for the susceptibility in terms
of QCD dressed quark propagator and dressed vector vertex. This
integral formula proves to reproduce the exact analytical result in
the case of a free quark gas. In the next section, with a model
gluon propagator input, the 2- and 3-point functions are confronted
with Dyson-Schwinger equation (DSE) and Bethe-Salpeter equation (BSE).
With these results, the quark number susceptibility is then
calculated and the characteristic behavior of the susceptibility
near the CEP is highlighted and its implications analyzed. The last
section is devoted to some concluding remarks. Throughout this work,
we work with Euclidean space metric: $\{\gamma_\mu,
\gamma_\nu\}=2\delta_{\mu\nu}$.

\section{ANALYTICAL RESULTS}
\subsection{Derivation of a closed integral expression for the quark number susceptibility}
The quark number susceptibility is defined as the derivative of
quark number density with respect to the baryon chemical potential
\begin{equation}
\chi=\frac{\partial n}{\partial\mu}.
\end{equation}
To start our derivation, we write the quark number density as
\begin{equation}
n=<\psi^+\psi>=<\overline{\psi}_i(\gamma_4)_{ij}\psi_j>=(-)N_cN_f\int\frac{d^4p}{(2\pi)^4}tr_{\gamma}[G(p,\mu)\gamma_4],
\end{equation}
where $G$ denotes the quark propagator, $N_c$, $N_f$ the color
factor and the number of flavors respectively, and the trace is taken
over Dirac indices. Here, the minus sign comes from the
anti-commutation property of the fermion fields.

Substituting Eq. (2) into Eq. (1) and adopting the identity
\begin{equation}
\frac{\partial G(p,\mu)}{\partial\mu}=(-)G(p,\mu)\frac{\partial
G^{-1}(p,\mu)}{\partial \mu}G(p,\mu),
\end{equation}
we have
\begin{equation}
\chi=N_cN_f\int\frac{d^4p}{(2\pi)^4}tr_{\gamma}[G(p,\mu)\frac{\partial
G^{-1}(p,\mu)}{\partial \mu}G(p,\mu)\gamma_4].
\end{equation}
Recall that the well-known Ward identity reads
\begin{equation}
i\Gamma_{\mu}(p,0)=\frac{\partial G^{-1}(p)}{\partial p_{\mu}},
\end{equation}
where $p$ denotes the relative momentum of the vector vertex and the
corresponding total momentum vanishes. Note that at finite
temperature and chemical potential, the fourth component of momentum
$p_4=\omega_n+i\mu$, with $\omega_n$ being fermion Matsubara
frequencies. Then
\begin{equation}
(-)\Gamma_4(p,0;\mu)=\frac{\partial G^{-1}(p,\mu)}{\partial \mu}.
\end{equation}
Putting this equation into Eq. (4) and replacing the integration over
the fourth component of momentum with explicit summation over
Matsubara frequencies, we get the quark number susceptibility at
finite temperature and chemical potential in the imaginary-time
formalism
\begin{equation}
\chi(T,\mu)=(-)N_cN_fT\sum_{n=-\infty}^{+\infty}\int\frac{d^3p}{(2\pi)^3}tr_{\gamma}[G(\widetilde{p}_n)\Gamma_4(\widetilde{p}_n,0)G(\widetilde{p}_n)\gamma_4],
\end{equation}
where $\widetilde{p}_n=(\overrightarrow{p},\omega_n+i\mu)$ with
fermion frequencies $\omega_n=(2n+1)\pi T$.

Thus a model-independent closed integral formula is obtained, which
expresses the quark number susceptibility in terms of the dressed quark
propagator and the dressed vector vertex, both of the latter objects
being basic quantities in quantum field theory. The DSE-BSE approach
provides an ideal framework to confront these quantities
non-perturbatively and then calculate the quark number
susceptibility.

\subsection{The free quark number susceptibility}
Before going into the model calculation of the quark number
susceptibility, let us apply the foregoing analytical expression for
the susceptibility to the case of a free quark gas. In this case,
the dressed vector vertex reduces to the bare one
\begin{equation}
\Gamma_4(\widetilde{p}_n,0)\longrightarrow\gamma_4
\end{equation}
and the free quark propagator reads
\begin{equation}
G^{free}(\widetilde{p}_n)=\frac{1}{i\gamma\cdot\widetilde{p}_n+m}
\end{equation}
with $m$ the degenerated light quark current mass. Substituting
Eqs. (8) and (9) into Eq. (7) and evaluating the trace, one gets
\begin{equation}
\chi^{free}(T,\mu)=4N_cN_fT\sum_{n=-\infty}^{+\infty}\int\frac{d^3p}{(2\pi)^3}\frac{(\omega_n+i\mu)^2-E^2_{\vec{p}}}{[(\omega_n+i\mu)^2+E^2_{\vec{p}}]^2},
\end{equation}
where $E_{\vec{p}}=\sqrt{\vec{p}^2+m^2}$. With
the aid of the identity
\begin{equation}
\sum_{n=-\infty}^{+\infty}\frac{1}{i\omega_n-x}=\frac{\beta}{2}\times\frac{1-e^{\beta
x}}{1+e^{\beta x}}
\end{equation}
one immediately arrives at
\begin{equation}
\chi^{free}(T,\mu)=2N_cN_f\beta\int\frac{d^3p}{(2\pi)^3} \left\{
\frac{e^{\beta(E_{\vec{p}}-\mu)}}{[e^{\beta(E_{\vec{p}}-\mu)}+1]^2}
+\frac{e^{\beta(E_{\vec{p}}+\mu)}}{[e^{\beta(E_{\vec{p}}+\mu)}+1]^2}
\right\}.
\end{equation}
This equation can be rewritten as
\begin{equation}
\chi^{free}(T,\mu)=2N_cN_f\frac{\partial}{\partial\mu}\int\frac{d^3p}{(2\pi)^3}\left\{\frac{1}{e^{\beta(E_{\vec{p}}-\mu)}+1}-\frac{1}{e^{\beta(E_{\vec{p}}+\mu)}+1}\right\}.
\end{equation}
This integral is nothing but the free quark gas density from the
Fermi-Dirac statistics. Eq. (13) is just a re-expression of the
definition of the quark number susceptibility (Eq.(1)) for a free
quark gas. The integration of Eq. (13) can be analytically solved in
the chiral limit ($m=0$)
\begin{equation}
\chi^{free}(T,\mu)=N_f(T^2+\frac{3\mu^2}{\pi^2}).
\end{equation}

We see that the integral formula Eq. (7) for the quark number
susceptibility reproduces exactly the Fermi-Dirac statistics result
for a free quark gas. This is to be expected in advance and can be
regarded as a consistent check of Eq. (7).

\section{MODEL CALCULATION}

\subsection{The dressed quark propagator}
In order to calculate the quark number susceptibility, we have to
evaluate first the dressed quark propagator and dressed vector
vertex at finite temperature and chemical potential. This will be
conducted systematically and consistently in the rainbow-DSE and
ladder-BSE \cite{a27} framework with a separable effective
interaction model.

The DSE-BSE provides a non-perturbative continuum approach for the
exploration of strong interaction physics. Its practical form,
namely the rainbow-ladder truncation, has given us fascinating
insights into the remarkable dynamical chiral symmetry breaking and
confinement \cite{a28,a29} and found successful applications in
describing light mesons in the pseudo-scalar and vector channels (for
review articles, see \cite{a30}). The rainbow DSE for the quark
propagator reads
\begin{equation}
G(p)^{-1}=i\gamma \cdot p
+m+\frac{4}{3}\int\frac{d^4p}{(2\pi)^4}g^2D_{\mu\nu}^{eff}(p-q)\gamma_{\mu}G(q)\gamma_{\nu}
\end{equation}
and the ladder inhomogeneous BSE for the dressed vector vertex reads
\begin{equation}
\Gamma_{\mu}(p,P)=\gamma_{\mu}-\frac{4}{3}\int\frac{d^4q}{(2\pi)^4}g^2D_{\rho\sigma}^{eff}(p-q)\gamma_{\rho}G(q_+)\Gamma_{\mu}(q,P)G(q_-)\gamma_{\sigma},
\end{equation}
where $P$ denotes the total momentum, $q_{\pm}=q\pm P/2$ and
$D_{\mu\nu}^{eff}(p-q)$ the effective gluon propagator, which is
usually a phenomenological input in practice (There are also some
attempts to explore the analytical structure of the gluon propagator
from numerical solutions of coupled DSEs of quarks, gluons and
ghosts and from fits to lattice date, see Ref. \cite{a29} and
references therein). In Refs. \cite{a31,a32}, the authors proposed a
confining, separable model gluon propagator with the rank-1 form
\begin{equation}
g^2D^{eff}_{\mu\nu}(p-q)=\delta_{\mu\nu}D_0f_0(p^2)f_0(q^2),
\end{equation}
where $f_0(p^2)=\exp(-p^2/\Lambda^2)$. This model is found to be
successful in describing light flavor pseudo-scalar and vector meson
observables with parameters $\Lambda_0=0.678~ \mathrm{GeV}$,
$D_0\Lambda_0^2=128.0$ and the degenerated light quark mass $m=6.6
~\mathrm{MeV}$ \cite{a32}.

In recent years, the rainbow-ladder DSE models were extended to
finite temperature and chemical potential to investigate QCD
thermodynamics (for a review, see \cite{a33}). As for the rank-1
confining separable model, the extension to finite temperature and
chemical potential is accomplished by transcription of the Euclidean
quark four-momentum via
$q\longrightarrow\widetilde{q}_n=(\vec{q},\widetilde{\omega}_n)$,
with $\widetilde{\omega}_n=\omega_n+i\mu$, and no new parameters are
introduced \cite{a32,a34,a35}. This means the effective gluon
propagator at finite temperature and chemical potential is modeled
as
\begin{equation}
g^2D^{eff}_{\mu\nu}(\widetilde{p}_k-\widetilde{q}_n)=\delta_{\mu\nu}D_0f_0(\widetilde{p}^2_k)f_0(\widetilde{q}^2_n).
\end{equation}
On the other hand, the finite temperature and chemical potential
version of the rainbow-DSE of quark propagator (Eq. 15) reads
\begin{equation}
G^{-1}(\widetilde{p}_k)=i\gamma\cdot\widetilde{p}_k+m+\frac{4}{3}T\sum_{n=-\infty}^{+\infty}
\int\frac{d^3q}{(2\pi)^3}g^2D^{eff}_{\mu\nu}(\widetilde{p}_k-\widetilde{q}_n)\gamma_{\mu}G(\widetilde{q}_n)\gamma_{\nu}
\end{equation}
and the quark propagator is generally decomposed as
\begin{equation}
G^{-1}(\widetilde{p}_k)=i\vec{\gamma}\cdot\vec{p}A(\widetilde{p}_k^2)+i\gamma_4\widetilde{\omega_k}C(\widetilde{p}_k^2)+B(\widetilde{p}_k^2).
\end{equation}
The rank-1 separable model leads to the rainbow-DSE solution:
$A(\widetilde{p}_k^2)=C(\widetilde{p}_k^2)=1$ and
$B(\widetilde{p}_k^2)=m+b(T,\mu)f_0(\widetilde{p}_k^2)$, where
$b(T,\mu)$ satisfies the following equation
\begin{equation}
b(T,\mu)=\frac{16}{3}D_0T\sum_{n=-\infty}^{+\infty}\int\frac{d^3q}{(2\pi)^3}\frac{f_0(\widetilde{q}_n^2)[m+b(T,\mu)f_0(\widetilde{q}_n^2)]}{[\widetilde{q}_n^2+(m+b(T,\mu)f_0(\widetilde{q}_n^2)]^2}.
\end{equation}
Of course, that the vector self-energy amplitudes $A$ and $C$ are
unity does not cover all the physical effects, but this respects the
nature of the separable model -- the correspondence between quark
propagator scalar self-energy amplitude $B$ and the separable gluon
propagator form factor $f_0$ \cite{a35}. Furthermore, it is an open
question how the chemical potential $\mu$ enters the gluon
propagator and the forgoing transcription of the fourth-component
momentum $p_4\longrightarrow\widetilde{\omega}_n=\omega_n+i\mu$ has
not been justified. However, this is again the requirement of the
nature of the separable model and we think it worthwhile to allow
such a uncomplicated extension to finite $T$ and $\mu$ in which the
Matsubara modes are not coupled and one could sum them as easily as
possible and then make a first survey of the thermodynamic
observables \cite{a35}. Fortunately the results turn out to be
satisfactory -- the correspondence nature of the separable model
plays a vital and correct role here \cite{a31}.

Eq. (21) is numerically solved. Note that this simple model quite
facilitates summation over fermion Matsubara frequencies as well as
the 3-momentum integration due to its ultraviolet finiteness
characterized by the Gaussian function $f_0(\widetilde{q}_n^2)$. As
a result, no renormalization is needed. The imaginary part of
$b(T,\mu)$ is found to be very small (of order $\sim 10^{-18}$) for
all temperatures and chemical potentials while the real part takes
strikingly different values at different $T$ and $\mu$ regions,
hence serving as the order parameter. Fig. 1 shows the
$\mu$-dependent behavior of $\mathrm{Re} b(T,\mu)$ at several fixed
temperatures. One sees that at low temperatures, $\mathrm{Re} b(T,\mu)$
demonstrates an abrupt discontinuous decrease to a very small value
at some critical chemical potential, which is typical of a phase
transition of first order and resembles the observations of
Nambu-Jona-Lasinio (NJL) model \cite{a36}. Nevertheless, at higher
temperatures, $\mathrm{Re} b(T,\mu)$ shows a smooth, gradual decrease as
$\mu$ increases, implying an analytic crossover. So one can
naturally expect a CEP at which the first-order phase transition and
the crossover connect. This will be verified by determining the most
singular point of the quark number susceptibility, which is
illustrated in subsection C. To summarize this subsection, we point
out that the observations of Fig. 1 agree with the scenario of QCD
phase transition shortly reviewed in the introduction.
\begin{figure}[ht]
   \includegraphics[width=12cm]{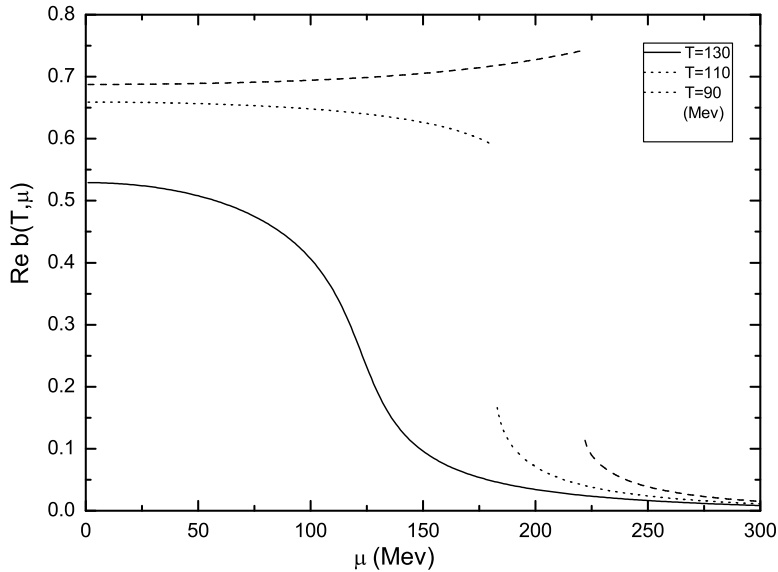}\\[-1cm]
  \caption{$\mu$-dependence of $\mathrm{Re} b(T,\mu)$ at three fixed temperatures.
  The evolution from a first-order phase transition to a crossover is shown.
  The different monotonous behavior of $\mathrm{Re} b(T,\mu)$ in the symmetry broken phase for
  $T=90~\mathrm{MeV}$ and $T=110~\mathrm{MeV}$ may be ascribed to the artifact of the separable model. }\label{fig1}
\end{figure}

\subsection{The dressed vector vertex}
We now turn to the dressed vector vertex. The finite temperature and
chemical potential version of the inhomogeneous ladder BSE for the
dressed vector vertex (the fourth component) Eq. (16) with vanishing
total momentum reads
\begin{equation}
\Gamma_4(\widetilde{p}_k,0)=\gamma_4-\frac{4}{3}T\sum_{n=-\infty}^{+\infty}\int\frac{d^3q}{(2\pi)^3}g^2D_{\rho\sigma}^{eff}(\widetilde{p}_k-\widetilde{q}_n)\gamma_{\rho}G(\widetilde{q}_n)\Gamma_4(\widetilde{q}_n,0)G(\widetilde{q}_n)\gamma_{\sigma}.
\end{equation}
where the model gluon propagator
$g^2D_{\rho\sigma}^{eff}(\widetilde{p}_k-\widetilde{q}_n)$ is given
by Eq. (18).

From Lorentz structure analysis, the dressed vector vertex with
vanishing total momentum at $T=0=\mu$ can be decomposed as
\cite{a30}
\begin{equation}
\Gamma_{\mu}(p,0)=\alpha_1(p^2)\gamma_{\mu}+\alpha_2(p^2)\gamma\cdot
pp_{\mu}-\alpha_3(p^2)ip_{\mu}.
\end{equation}
This can also be easily seen from the Ward identity
$i\Gamma(p,0)=\partial G^{-1}(p)/\partial p_{\mu}$ with
$G^{-1}(p)=i\gamma\cdot pA(p^2)+B(p^2)$. In Eq. (23), $\alpha_1(p^2)$
is the dominant term, while the other two terms contribute so little
that they can even be neglected \cite{a30}. At finite temperature
and chemical potential, we keep the corresponding "leading" term in
$\Gamma_4(\widetilde{p}_k,0)$ and add another term
\begin{equation}
\Gamma_4(\widetilde{p}_k,0)=\alpha(\widetilde{p}_k^2)\gamma_4-i\beta(\widetilde{p}_k^2)\widetilde{\omega}_k.
\end{equation}
Here we stress that the second term is vital and indispensable for
describing the QCD phase transition at finite temperature and
chemical potential. Some explanation goes as follows in this
connection. From the Ward identity Eq. (6) one sees that this term
arises from the derivative of the scalar self-energy amplitude
$B(\widetilde{p}_k^2)$ of the inverse dressed quark propagator
Eq. (20) with respect to the chemical potential $\mu$. Note that
$B(\widetilde{p}_k^2)$ experiences an abrupt decrease at some
critical $\mu$ when a true phase transition occurs (on the contrary,
the vector self-energy amplitudes $A(\widetilde{p}_k^2)$ and
$C(\widetilde{p}_k^2)$ evolve smoothly with $T$ and $\mu$
\cite{a32}) and hence makes a considerable contribution to
$\Gamma_4(\widetilde{p}_k,0)$ at that point.

Putting the ansatz for the dressed vector vertex Eq. (24), the
dressed quark propagator Eq. (20) and the confining separable model
gluon propagator Eq. (18) into Eq. (22), one can transform the ladder BSE
into two coupled non-linear equations
\begin{eqnarray}
u=\frac{8}{3}D_0T\sum_{l=-\infty}^{+\infty}\int\frac{d^3q}{(2\pi)^3}f_0(\widetilde{q}_l^2)
\frac{[1-uf_0(\widetilde{q}_l^2)][\widetilde{\omega}_l^2-\vec{q}^2-(m+bf_0(\widetilde{q}_l^2))^2]+2vf_0(\widetilde{q}_l^2)\widetilde{\omega}_l[m+bf_0(\widetilde{q}_l^2)]}
{[\widetilde{q}_l^2+(m+bf_0(\widetilde{q}_l^2))^2]^2}
\\
v=\frac{16}{3}D_0T\sum_{l=-\infty}^{+\infty}\int\frac{d^3q}{(2\pi)^3}f_0(\widetilde{q}_l^2)
\frac{vf_0(\widetilde{q}_l^2)[\widetilde{q}_l^2-(m+bf_0(\widetilde{q}_l^2))^2]-2[1-uf_0(\widetilde{q}_l^2)]\widetilde{\omega}_l[m+bf_0(\widetilde{q}_l^2)]}
{[\widetilde{q}_l^2+(m+bf_0(\widetilde{q}_l^2))^2]^2}
\end{eqnarray}
with $\alpha(\widetilde{p}_k^2)=1-u(T,\mu)f_0(\widetilde{p}_k^2)$
and
$\beta(\widetilde{p}_k^2)\widetilde{\omega}_k=v(T,\mu)f_0(\widetilde{p}_k^2)$.
With $b(T,\mu)$ obtained from Eq. (21), these two equations are
numerically solved. Here, one also notes that the Gaussian function
$f_0(\widetilde{q}_l^2)$ appearing in the numerators of the
integrands justifies a numerical cut-off for summation over
Matsubara frequencies as well as 3-D momentum integration.

\subsection{Results for the quark number susceptibility}
Having completed the calculation of the dressed quark propagator and
the dressed vector vertex, we are now in a position to calculate the
quark number susceptibility. The numerical results are shown in
Fig. 2, where $\chi(T,\mu)$ is normalized by the free quark quark
number susceptibility in the chiral limit
$\chi^{free}(T,\mu)$ (Eq. (14)) and hence dimensionless.
\begin{figure}[ht]
   \includegraphics[width=12cm]{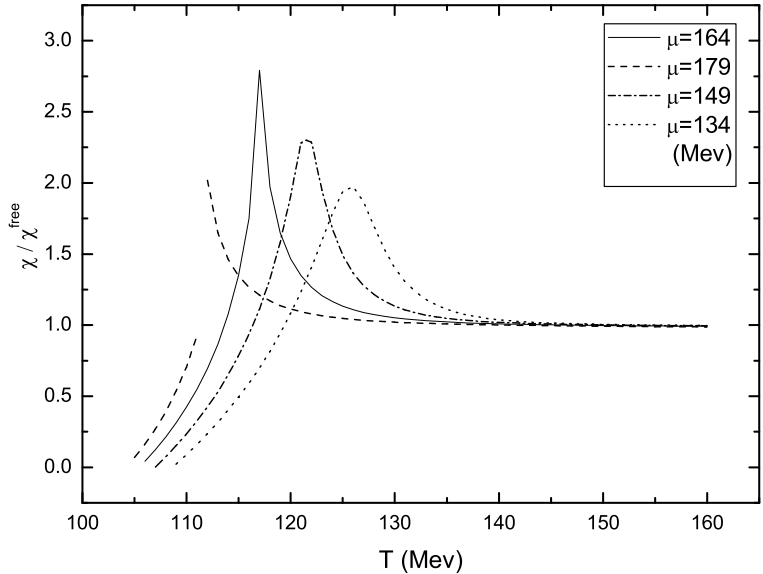}\\[-1cm]
  \caption{$T$-dependence of $\chi(T,\mu)$ at four fixed chemical potentials. From left to right, $\mu=179, 164, 149, 134~\mathrm{MeV}$,
   respectively. The CEP is identified as the diverging cusp developed in the solid line at $T_{\mathrm{CEP}}=117~\mathrm{MeV}, \mu_{\mathrm{CEP}}=164~\mathrm{MeV}$.}\label{fig2}
\end{figure}

Let us analyse the observations from Fig. 2. The most singular point
develops along the $\mu_{\mathrm{CEP}}=164~\mathrm{MeV}$ solid line
at $T_{\mathrm{CEP}}=117~\mathrm{MeV}$. This is the CEP, at which
the quark number susceptibility manifests itself as a diverging cusp
that should translate into event by event fluctuations of baryon
multiplicity \cite{a22}. For larger $\mu=179~\mathrm{MeV}$ (the
dashed line), $\chi(T,\mu)$ also jumps across the chiral phase
transition and has a discontinuity due to its first-order character.
The peak is still pronounced but its height much suppressed. For
smaller chemical potentials ($\mu=149, 134~\mathrm{MeV}$, the
dash-dotted and the dotted line), one enters a crossover region, in
which the discontinuity of $\chi(T,\mu)$ at the transition vanishes
and $\chi(T,\mu)$ evolves gradually with $T$ in a smoother and
smoother manner.

Another observation from Fig. 2 is that far below the chiral phase
transition, $\chi(T,\mu)$ is greatly suppressed and at asymptotic
temperatures, $\chi(T,\mu)$ tends unambiguously to the value of
susceptibility of a free quark gas. In the latter respect, our
demonstration is more akin to that of lattice predictions
\cite{a23,a25} than NJL-type models \cite{a18,a20}, where $\chi$
decreases too fast at asymptotic temperatures or chemical
potentials. This may be accounted for by the fact that the
momentum-dependent interaction DSE model represents a better
asymptotic freedom behavior.

Our calculation determines the CEP location at $T_{\mathrm{CEP}}=117~\mathrm{MeV}$ and
$\mu_{\mathrm{CEP}}=164~\mathrm{MeV}$. The most recent experimental estimate by
extracting $\eta/s$ as a function of $T$ and $\mu$ from an elliptic
flow excitation function gives $T_{\mathrm{CEP}}\sim165-170~\mathrm{MeV}$ and
$\mu_{\mathrm{CEP}}\sim150-180~\mathrm{MeV}$ \cite{a16}, which is in good agreement
with lattice prediction with almost realistic pion mass \cite{a24}.
So very encouragingly, our results for $\mu_{\mathrm{CEP}}$ lies exactly in
this interval, whereas our $T_{\mathrm{CEP}}$ deviates a lot, for the
pseudo-critical temperature $T_c$ in our calculation takes a relatively lower
value (see below). As for NJL-type models, they predict a much higher
$\mu_{\mathrm{CEP}}$ \cite{a18,a20,a22}. In fact, NJL-type models have a
common tendency to lead to the CEP at relatively high
$\mu_{\mathrm{CEP}}\sim 300~\mathrm{MeV}$ \cite{a22}.

Having identified the CEP, we now show the complete phase diagram obtained
from our model, which assumes the shape sketched in Fig. 3.
\begin{figure}[ht]
   \includegraphics[width=12cm]{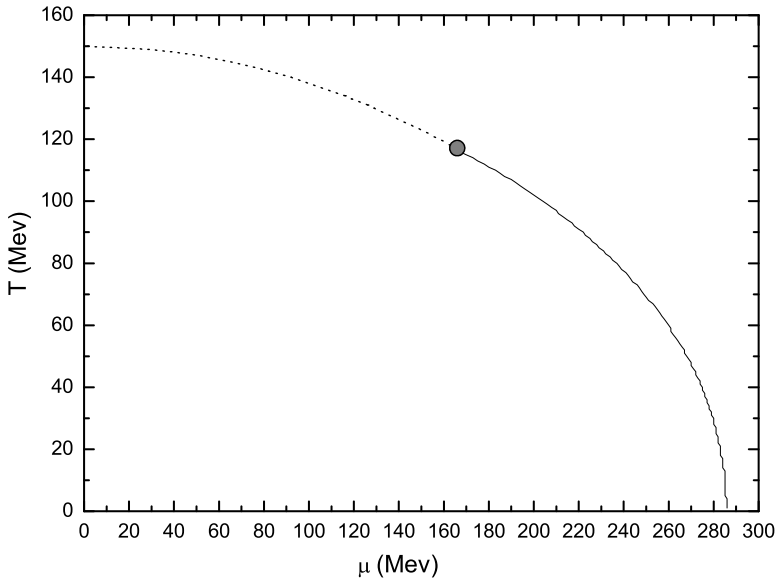}\\[-1cm]
  \caption{The phase diagram characterized by $\mathrm{Re} b(T,\mu)$. The solid line represents
  the first-order phase transition and the dotted line the crossover transition. The filled
  circle indicates the CEP.}\label{fig3}
\end{figure}
Two separate phases are shown in the $T-\mu$ plane: the chiral
symmetry restored phase and the phase with (approximate) chiral
symmetry dynamically broken. The solid line represents a first-order
phase transition line, which begins from $\mu_c=286~\mathrm{MeV}$ at vanishing
temperature and ends at the CEP ($T_{\mathrm{CEP}}=117~\mathrm{MeV}, \mu_{\mathrm{CEP}}=164~\mathrm{MeV}$).
Beyond the $T_{\mathrm{CEP}}$, one ends up with a crossover (the dotted line)
with a pseudo-critical temperature $T_c=150~\mathrm{MeV}$ at vanishing
chemical potential.

Last but not least, to check the parameter sensitivity of the
location of the critical end point, we first vary the degenerated
current quark mass $m$, since the value of $m$ is believed to be
closely associated with the nature of the phase transition. The
numerical results is shown in Fig. 4.

\begin{figure}[ht]
   \includegraphics[width=12cm]{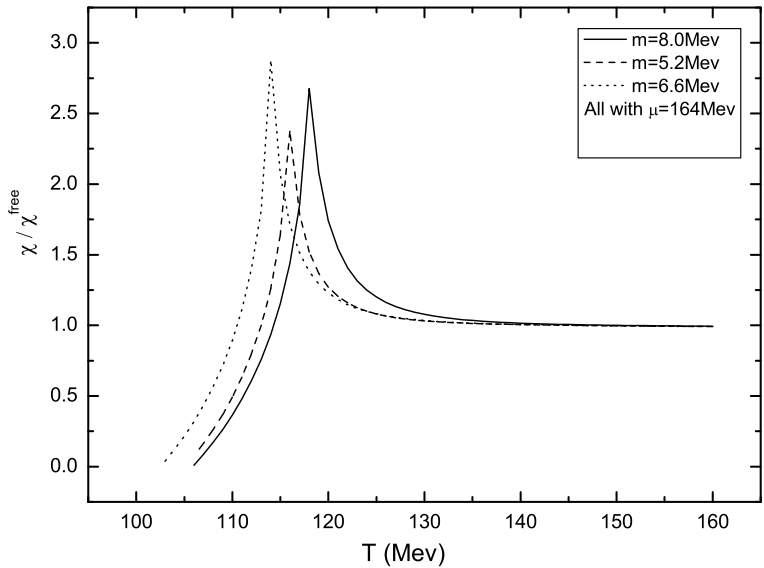}\\[-1cm]
  \caption{The location of the critical end point for $m=5.2~\mathrm{MeV}$ and $m=8.0~\mathrm{MeV}$ with "physical" $D_0=128.0/(0.687 ~\mathrm{GeV})^2$ and for "physical" $m=6.6~\mathrm{MeV}$ with $D_0=128.0/(0.700~\mathrm{GeV})^2$. For all three curves, the chemical potential takes the same value
$\mu_{\mathrm{CEP}}=164~\mathrm{MeV}$.}\label{fig4}
\end{figure}

We find that when the quark mass value varies from $m=5.2~\mathrm{MeV}$ to
$m=8.0~\mathrm{MeV}$, the most singular point of the quark number
susceptibility emerges always along the $\mu_{\mathrm{CEP}}=164~\mathrm{MeV}$ line,
while the corresponding $T_{\mathrm{CEP}}$ has a very minor shift from
$116~\mathrm{MeV}$ to $118~\mathrm{MeV}$ (the dashed and solid line, respectively). As we
change the coupling parameter from its physical value in the present
model $D_0=128.0/(0.687~\mathrm{GeV})^2$ to $D_0=128.0/(0.700~\mathrm{GeV})^2$ (with
quark mass fixed at its physical value), the corresponding CEP moves
slightly to $T_{\mathrm{CEP}}=114~\mathrm{MeV}$, with its $\mu_{\mathrm{CEP}}$ remaining at
$164~\mathrm{MeV}$. Lastly, as for the parameter $\Lambda$, the effect of its
variation can be canceled by taking a larger numerical cutoff in
practical calculations, since it appears in the exponential as the
denominator. Also notable here is that the critical line does not change
much, either for $m=5.2~\mathrm{MeV}$, the critical chemical potential at
vanishing temperature is $\mu_c=280~\mathrm{MeV}$, for $m=8.0~\mathrm{MeV}$,
$\mu_c=289~\mathrm{MeV}$, and for the new coupling parameter
$D_0=128.0/(0.700~\mathrm{GeV})^2$, $\mu_c=281~\mathrm{MeV}$, while all the
pseudo-critical temperatures take almost the same value as that for
the ``physical'' parameters.

The conclusion is that the location of the CEP (especially 
$\mu_{\mathrm{CEP}}$) and the critical line is relatively robust with respect
to the variations of the model parameters around their physical
values.

\section{Summary and conclusions}
The primary goal of this paper is to identify and locate the CEP
through a continuum study of the quark number susceptibility. To
this end, we first derive a model-independent integral formula
expressing the quark number susceptibility in terms of dressed
quark propagator and dressed vector vertex. The latter objects are
then confronted in the rainbow-ladder DSE-BSE framework with a
confining separable model gluon propagator. With these results, the
quark number susceptibility is calculated. The quark number
susceptibility measures the response of the quark number density to
changes in the baryon chemical potential and is of particular
interest for exploring the CEP in that it is expected to diverge at
the CEP. Our calculation reproduces this diverging phenomenon, which
may be detectable in heavy ion collision experiments \cite{a10,a25}.

The location of the CEP our calculation predicts lies at a
reasonable value. Interestingly the $\mu_{\mathrm{CEP}}$ is
coincident with recent experimental estimate and more acceptable
than NJL-type model predictions. In this regard, we emphasize that
the confining separable DSE model admits a momentum-dependent
effective interaction, which is not the case for NJL-type models.
Moreover, the rainbow-ladder truncation scheme for DSE-BSE respects
the axial-vector Ward-Takahashi identity (AV-WTI) \cite{a30}, which
relates closely to the chiral symmetry and its dynamical breaking,
and the usual Abelian WTI, which plays an essential part in our
analysis in this paper.

Nevertheless, our model study says nothing about the recently
prevalent terminology in QCD thermodynamics, namely the strongly
coupled Quark-Gluon Plasma(sQGP) \cite{a37}. $\eta/s$, rather than
the quark number susceptibility, is believed to provide a good
characteristic of sQGP liquid and its surprisingly small value makes
a current focus \cite{a15,a37}, with which we are also concerned.

\section{Acknowledgment}
We acknowledge helpful communications with P. C. Tandy and D.
Blaschke on the confining separable model. This work was supported
in part by the National Natural Science Foundation of China (under
Grant Nos 10575050 and 10775069) and the Research Fund for the
Doctoral Program of Higher Education (under Grant No 20060284020).


\begin{references}


\bibitem{a1}M. Stephanov, arXiv:hep-lat/0701002v1 (2006).
\bibitem{a2}U. Heinz, arXiv:hep-ph/0407360v1 (2004).
\bibitem{a3}M. He, D. K. He, H. T. Feng, W. M. Sun, and H. S. Zong, Phys. Rev. {\bf D 76}, 076005 (2007).
\bibitem{a4}F. Wilczek, Nature {\bf443}, 637 (2006).
\bibitem{a5}Y. Aoki et al., Nature {\bf443}, 675 (2006).
\bibitem{a6}Y. Aoki et al., Phys. Lett. {\bf B 643}, 46 (2006).
\bibitem{a7}F. Karsch, arXiv:hep-lat/0711.0656v1 (2007).
\bibitem{a8}O. Philipsen, arXiv:hep-lat/0708.1293v2 (2007).
\bibitem{a9}M. Stephanov, Prog. Theor. Phys. Suppl. {\bf153}, 139(2004); Int. J. Mod. Phys. {\bf A 20}, 4387
(2005).
\bibitem{a10}M. Stephanov, K. Rajagopal, and E. Shuryak, Phys. Rev. Lett. {\bf 81,} 4816
(1998); Phys. Rev. {\bf D 60}, 114028 (1999).
\bibitem{a11}Y. Hatta and M. A. Stephanov, Phys. Rev. Lett. {\bf91},
102003 (2003).
\bibitem{a12}Z. Fodor and S. D. Katz, JHEP {\bf04}, 050 (2004); Shinji
Ejiri, Phys. Rev. {\bf D 77}, 014508 (2008).
\bibitem{a13}M. A. Stephanov, Phys. Rev. {\bf D 73}, 094508 (2006).
\bibitem{a14}B. Lungwitz and M. Bleicher, Phys. Rev. {\bf C 76}, 044904
(2007).
\bibitem{a15}R. A. Lacey et al., Phys. Rev. Lett. {\bf98}, 092301
(2007).
\bibitem{a16}R. A. Lacey et al., arXiv:nucl-ex/0708.3512v6 (2008).
\bibitem{a17}Yoshitaka Hatta and Takashi Ikeda, Phys. Rev. {\bf D 67}, 014028
(2003).
\bibitem{a18}P. Costa, C. A. de Sousa, M. C. Ruivo and Yu. L. Kalinovsky, Phys. Lett. {\bf B 647}, 431 (2007); P. Costa, M. C. Ruivo, C. A. de Sousa, Phys. Rev. {\bf D 77}, 096001 (2008).
\bibitem{a19}C. Sasaki, B. Friman, and K. Redlich, Phys. Rev. {\bf D 75}, 074013 (2007).
\bibitem{a20}B.-J. Schaefer and J. Wambach, Phys. Rev. {\bf D 75}, 085015 (2007).
\bibitem{a21}K. Redlich, B. Friman, and C. Sasaki, J. Phys. G: Nucl. Part. Phys.
{\bf 35} 044013 (2008).
\bibitem{a22}Kenji Fukushima, Phys. Rev. {\bf D 77}, 114028 (2008).
\bibitem{a23}C. R. Allton et al., Phys. Rev. {\bf D 71}, 054508 (2005).
\bibitem{a24}R.V. Gavai and Sourendu Gupta, Phys. Rev. {\bf D 71}, 114014
(2005).
\bibitem{a25}C. Schmidt, arXiv:hep-lat/0610116v2 (2006); O. Philipsen, arXiv:hep-lat/0510077 (2005).
\bibitem{a26}L. P. Csernai, J. I. Kapusta, and L. D. McLerran, Phys.
Rev. Lett. {\bf97}, 152303 (2006); J.-W. Chen and E. Nakano,
arXiv:hep-ph/0604138 (2006).
\bibitem{a27}C. D. Roberts, and A. G. Williams, Prog. Part. Nucl. Phys. {\bf33}, 477 (1994).
\bibitem{a28}R. Alkofer and L. von Smekal, Phys. Rep. {\bf 353}, 281
(2001).
\bibitem{a29}C. S. Fischer, J. Phys. {\bf G 32}, R253 (2006).
\bibitem{a30}P. Maris and C. D. Roberts, Int. J. Mod. Phys. {\bf E 12}, 297
(2003); A. Holl, C. D. Roberts, and S. V. Wright, nuch-th/0601071 v1
(2006).
\bibitem{a31}C. J. Burden, L. Qian, C. D. Roberts, P. C. Tandy, and M. J. Thomson, Phys. Rev. {\bf C 55}, 2649 (1997).
\bibitem{a32}D. Blaschke et al., Int. J. Mod. Phys. {\bf A 16}, 2267 (2001).
\bibitem{a33}C. D. Roberts and S. M. Schmidt, Prog. Part. Nucl. Phys. {\bf45}, S1 (2000).
\bibitem{a34}D. Blaschke and P.C. Tandy, arXiv:nucl-th/9905067v1
(1999); D.Blaschke and C. D. Roberts, Nucl. Phys. {\bf A 642}, 197
(1998).
\bibitem{a35}Private communications with P. C. Tandy and D. Blaschke.
\bibitem{a36}M. Buballa, Phys. Rep. {\bf407}, 205 (2005).
\bibitem{a37}E. Shuryak, Nucl. Phys. {\bf A 774}, 387 (2006);
arXiv:hep-ph/0804.1373v1 (2008).




\end{references}
\end{document}